\documentclass[prd,twocolumn,superscriptaddress,showpacs,nofootinbib,preprintnumbers]{revtex4}

\usepackage{graphicx,bm}

\usepackage{dcolumn}    

\begin{document}

\title{Are black holes over-produced during preheating?}
\author{Teruaki Suyama}
\email{suyama_at_tap.scphys.kyoto-u.ac.jp}
\affiliation{Department of Physics, Kyoto University, Kyoto 606-8502, Japan}

\author{Takahiro Tanaka}
\email{tama_at_scphys.kyoto-u.ac.jp}
\affiliation{Department of Physics, Kyoto University, Kyoto 606-8502, Japan}

\author{Bruce Bassett}
\email{Bruce.Bassett_at_port.ac.uk}
\affiliation{Department of Physics, Kyoto University, Kyoto 606-8502, Japan}
\affiliation{Institute of Cosmology and Gravitation, University of Portsmouth, Mercantile House, Portsmouth PO1 2EG, UK}

\author{Hideaki Kudoh}
\affiliation{Department of Physics, The University of Tokyo, Bunkyo-ku, 113-0033, Japan}
\email{kudoh_at_utap.phys.s.u-tokyo.ac.jp}
\preprint{KUNS-1939, UTAP-498}

\date{\today}

\begin{abstract}
We provide a simple but robust argument that primordial black hole (PBH) production generically does {\em not} exceed 
astrophysical bounds during the resonant preheating phase after inflation. This conclusion is supported by  
fully nonlinear lattice simulations of various models in two and three dimensions which include rescattering but neglect metric 
perturbations.  We examine the degree to which preheating amplifies density perturbations at the Hubble scale and show that 
at the end of the parametric resonance, power spectra are universal, with no memory of the power spectrum 
at the end of inflation. In addition we show how the probability distribution of density perturbations changes from exponential on very 
small scales to Gaussian when smoothed over the Hubble scale --  the crucial length for studies of primordial black hole formation -- hence 
justifying the standard assumption of Gaussianity.
\end{abstract}

\pacs{98.80.Cq, 97.60.Lf}

\maketitle

\section{Introduction}
Primordial black holes (PBHs) span a wide range of mass scales and are typically much smaller
than the solar mass ($ \sim 10^{33} {\rm g} $) and may be formed in the early universe~\cite{carr}. 
PBHs may form from  the gravitational collapse of large density fluctuations at horizon (i.e. Hubble scale $k = aH$) 
crossing  in the radiation dominated universe.  A PBH formed at the Planck time 
$ \sim 10^{-43} {\rm sec} $ will have a mass $ \sim 10^{19} {\rm GeV} $, while masses around 
$ \sim 10^{15} {\rm g} $ are formed at $ \sim 10^{-23} {\rm sec} $ ( see for example~\cite{carr2} ).
 
The evaporation time for a PBH mass $ \sim 10^{15} {\rm g} $ is nearly the present age of the universe, 
so PBH with smaller masses than this would have evaporated in the past, unloading a potentially vast amount 
of entropy. The success of the standard cosmology and the observation of the cosmic rays 
severely constrains the abundance of PBHs for various masses and provides useful constraints on inflationary and 
early universe physics. 

For example, Big Bang Nucleosynthesis limits the PBH abundance in the mass range $ 10^{6} \sim 10^{13} {\rm g} $, 
by limiting the entropy from  Hawking radiation or requiring that it does not modify the cosmological composition 
of the light elements~\cite{starobinskii}. PBH of mass $ 10^{15} {\rm g} $ evaporating now will emit particles such 
as $ \gamma$-rays which are constrained by observations of the extragalactic $\gamma$-ray background which imply
$ {\Omega}_{PBH}  < 10^{-8} $~\cite{page}.
For $ > 10^{15} {\rm g} $, the limit on the PBH abundance is obtained from requiring that $ {\Omega}_{PBH} $ 
does not exceed unity. 

These observational constraints on PBHs provide a powerful probe of the primordial fluctuations.
The upper bound on the abundance of PBHs directly leads to that on the density fluctuation
at horizon crossing when PBH are formed. Therefore the scales of fluctuations relevant to 
PBH formation are much smaller than those associated with the Cosmic Microwave Background and 
the large-scale structure. This makes studying PBHs important. In the past, for example,
constraints on the density perturbation spectrum were obtained by studying PBH formation~\cite{bringmann}.
While the requirement that PBHs are not over-produced yields useful information about
the early universe, PBH can be an interesting dark matter candidate in smaller abundances~\cite{ivanov}.

In this paper we will show that typically PBH are not over-produced during the violent non-equilibrium phase of preheating that
follows the end of many inflationary models. This follows from three key observations: (1) the peak of the density perturbation spectrum
typically lies at scales smaller than the Hubble scale. (2) The peak corresponds to density contrasts of order unity. (3) The slope of the spectrum 
around the horizon size is three (in $3d$). Putting these together we typically find that at the horizon scale relevant for PBH formation, the 
density contrast is around an order of magnitude too small to over-produce PBH.  Nevertheless, the density perturbation on the horizon scale 
is significantly enhanced by preheating (by several orders of magnitude) compared with the no-resonance case and hence preheating 
is important in understanding the potential astrophysical and cosmological implications of PBH. 

In studying the production of PBHs, one usually assumes that the probability distribution of 
density fluctuations at horizon crossing is Gaussian. This assumption is critical 
because the density perturbations which collapse to black holes are very rare: several $ \sigma $ 
fluctuations (otherwise PBH will be over-produced in all cases) and therefore production of PBHs is
sensitive to the tail of the distribution. Indeed Bullock and Primack~\cite{bullock}
found that in some inflation models large perturbations are suppressed relative to 
a Gaussian distribution, resulting in a significant change in a number of PBH.
We study the validity of the Gaussian assumption in a later section.

Among various possible scenarios that might over-produce PBHs, we focus on 
preheating after inflation. Preheating is a process in which energy transfer occurs
rapidly from inflaton field to another field due to the non-perturbative effects during 
the oscillating phase of the inflaton~\cite{traschen}. This process significantly differs from the usual 
reheating scenario, where inflaton decays perturbatively to another particles, in a sense 
that in preheating most energies of inflaton field converts to created particles only
during the several oscillations of inflaton and even the massive particle which is 
much heavier than the inflaton can be created. It has been understood that parametric 
resonance occurs generically at the first stage of reheating~\cite{kofman1}. The parametric 
resonance does not last long because the rapid increase of created particles eventually
affect the motion of background field and the created particles scatter off each other, 
removing the particles from the resonance band. By these effects, the resonance becomes
inefficient and the decaying process of inflaton is described by usual single-body decay
theory which finally leads to the thermal equilibrium state~\cite{felder00}.

There are many works on preheating. The first stage of preheating, where the backreaction is
negligible and the linear approximation is valid, was studied in detail in~\cite{kofman2, pgreene, shtanov}.
Parametric resonance including  metric perturbations in order to see the behavior on super horizon scales were studied 
in~\cite{taruya, kodama, bassett1, bassett2, finelli1, jedamzik, liddle}. As in the case without metric perturbations, there is
a crucial difference between the single and multiple field cases. The analysis of fully non-linear 
preheating including gravity is very difficult. Before now, different approximations 
such as mean-field approximation~\cite{bassett2, jedamzik, zibin, bassett3, bassett4,tsujikawa, boyanovsky}, 
lattice simulation~\cite{lattice} without metric perturbation and 
one dimensional fully non-linear calculations~\cite{finelli2, parry, easther} have been done for studying
the various effects caused by preheating on the present universe.

Green and Malik~\cite{agreen} argued that PBHs {\em will} be overproduced due to the amplification of fluctuations 
during preheating for many parameter regions in a two-field massive inflation model based on
the results of~\cite{liddle2}, which takes into account of the second order fluctuation of $ \chi $ field. Put simply, their 
results suggested that the backreaction timescale was smaller than the timescale for the over-production of PBH. 

Bassett and Tsujikawa~\cite{bassett3} studied PBH production in the two-field massless inflation model including 
the effect of backreaction via the Hartree-Fock approximation  and found that 
PBH overproduction might occur, if the  probability distribution of fluctuation at horizon crossing was
assumed to be chi-squared (which lowers the threshold mass variance, $\sigma$). 
However, they found that PBH were not over-produced if the distribution was assumed to be Gaussian 
{\em and} the density field was smoothed on the scale of the horizon. Nevertheless, that analysis was limited since it 
neglected the mode-mode coupling  effects of rescattering and hence there was an open question 
both as to the underlying probability distribution of the density fluctuations
and the contribution of rescattering to horizon-scale fluctuations. 

We address both of these issues in this paper. We study PBH production
due to preheating via two and three-dimensional lattice simulations which
automatically include the effects of backreaction and rescatterings. 
We modified the C++ code LATTICEEASY written by Felder and Trachev~\cite{latticeeasy}.
In lattice simulations, the evolution equations for the scalar fields (and also the scale factor) 
are solved in real (as opposed to Fourier) space
($ N=2048^2 $ for 2 dimensional simulation and $ N=128^3 $ for 3 dimensional simulation). 
Metric perturbations were not included, so we cannot 
apply this method to the dynamics on super horizon scales. We followed 
the evolution of both the scalar fields and the total density perturbation. We found that PBH are not overproduced
and, interestingly, that the power spectrum at the end of preheating has a universal feature, that is, it is determined 
by the preheating dynamics and does not depend on the initial conditions. We also studied the 
probability distribution of the density perturbation at horizon crossing and found that it 
remained Gaussian which seems to be valid even in the tail of distribution.

A brief summary of our paper is as follows.
Section II gives a brief review of calculating the PBH abundance formed from the large 
density perturbation. Section III describes the model of preheating in this simulation.
In section IV, we consider the initial conditions of scalar fields. Section V shows 
the numerical results of lattice simulation. We will interpret the non-linear behavior 
of fluctuations and study the production of PBHs. Section VI discusses the probability 
distribution of fluctuations at horizon crossing we assume Gaussian in usual 
and Section VII is a conclusion.

\section{PBH formation by parametric resonance}
\subsection{Abundance of PBHs}
In this subsection, we briefly review the standard method 
to estimate the abundance of PBHs~\cite{carr3}. 
In the radiation dominated universe
PBHs will be produced 
if $ {\delta}_H $, the amplitude of density perturbations $ \delta $ smoothed over 
the horizon size in the comoving gauge, 
exceeds a certain threshold ${\delta}_c$~\cite{agreen2, harrison}. 
From linear analysis the critical value of $ {\delta}_c $ is roughly estimated to be 
$1/3$, but it is not independent of the initial density profile. 
Numerical study~\cite{niemeyer} suggest 
$\delta_c\sim 0.7 $ 
for various initial density profiles(see also ~\cite{shibata}). 

Under the assumption that the probability distribution 
of density fluctuation at horizon crossing is 
Gaussian,
the mass fraction of PBHs at the formation time, 
$\beta $, is estimated as 
\begin{equation}
\beta = {\int}_{ {\delta}_c }^{\infty} d {\delta}_H \ P ( {\delta}_H ) \sim \frac{\sigma}{ \sqrt{2 \pi} {\delta}_c }
\exp \left( - \frac{ {\delta}^2_c}{2 {\sigma}^2} \right), \label{pbh2}
\end{equation}
where $\sigma^2$ is the variance of $\delta_H$. 
Since the mass fraction of PBHs increases in proportion to 
the scale factor in the radiation dominated 
universe, $ \beta $ must be very small in order to 
satisfy the astrophysical constraints.

Roughly speaking, $ \beta $ is 
observationally constrained to be smaller
than about $ 10^{-20}$ for most of the range of PBH mass except 
for a small window at $M\approx 10^{15}$g, which corresponds to 
PBHs evaporating now. 
If we adopt $ 10^{-20} $ as the upper bound on $ \beta $, 
the upper bound on $ \sigma $ becomes $ \sim 0.03 $ and 
$ \sim 0.08 $ for $ {\delta}_c = 0.3 $ and $ 0.7 $, respectively, assuming a Gaussian 
distribution for $\delta$.
Though there is uncertainty in $ {\delta}_c $  it does not affect our conclusions in
the range of values $ 0.3 - 0.7 $. 
Therefore we adopt the smaller value $ 0.03 $ as 
the upper bound on $ \sigma $ to be conservative.

To estimate the variance of density perturbations at the horizon
scale, we simply use 
the power spectrum ${\cal P}_{\delta} (aH)$, 
where $ a $ and $H$ are, respectively, 
the scale factor and the Hubble parameter,
and ${\cal P}_{\delta} (k)$ is defined by
\begin{equation}
\langle {\delta}_k {\delta}_{-k'} \rangle = \frac{2 {\pi}^2}{k^3} {\cal
P}_{\delta} (k) \delta ( \vec{k}-\vec{k'}). \label{pbh5} 
\end{equation}

\subsection{Models of parametric resonance}
In this paper, we consider two simple models of preheating. 
\subsubsection{Conformal Models}
Conformal models are models composed of two scalar field 
with the potential given by~\cite{pgreene},
\begin{equation}
V(\phi, \chi) = \frac{\lambda}{4} {\phi}^4 + \frac{g^2}{2} {\phi}^2 {\chi}^2. 
\label{model1}
\end{equation}
Here $ \phi $ is the inflaton field. 
We start our simulation at the time when $\phi$ drops down to 
$0.34 m_{pl}$ \cite{pgreene}, where $m^2_{pl}=1/G$. 
In the oscillating phase of the inflaton 
the universe is effectively radiation dominated 
for the potential quartic in fields when averaged in time. 

As standard, we introduce rescaled fields by 
\begin{equation}
{\tilde \phi} = a \phi/\phi_0, \ \ \ {\tilde \chi} = a \chi/\phi_0,
\label{model2} 
\end{equation}
where the scale factor $a$ is normalized to unity
at the end of inflation, ({\it i.e.}, at the beginning of preheating), 








We also introduce the rescaled conformal time $\tilde \eta$,  
related to proper time $t$ by 
\begin{equation}
 a d\tilde \eta=\sqrt{\lambda} \phi_0 dt.
\end{equation}
Then, the equations of motion for this model become
\begin{eqnarray}
{\tilde \phi}'' - \lambda^{-1}\triangle {\tilde \phi} +
{\tilde \phi}^3 + {g^2\over\lambda} {\tilde \phi} {\tilde \chi}^2 
- \frac{a''}{a} {\tilde \phi}=0, 
\label{model91}
\\
{\tilde \chi}'' - \lambda^{-1}\triangle {\tilde \chi} + 
{g^2\over\lambda} {\tilde \phi}^2 {\tilde \chi} 
- \frac{a''}{a} {\tilde \chi}=0,  
\label{model92}
\end{eqnarray}
where $ ' $ denotes the differentiation with respect to $\tilde\eta$.
In the radiation dominated universe ( $ a \propto \tilde\eta $) 
the last terms in Eqs.~(\ref{model91}) and (\ref{model92}) 
proportional to $ a'' $ vanish, 
and thus these equation reduce
to the Minkowski ones. This is only exactly true if $\phi$ were conformally coupled 
to the curvature but since this is a weak effect we neglect it.

The unperturbed background solution for ${\tilde \phi}$ is given 
by Jacobi's elliptic cosine function, 
$\mbox{cn} (\tilde\eta)$ \cite{pgreene}. 
Then, linearized equations obey the so-called Lam\'e
equation~\cite{finkel} with 
resonance parameters $ 3 $ and $g^2/\lambda$ for $\phi$ and $\chi$, 
respectively. 
Hence, the growth rate of the longest wavelength mode for $\chi$ 
is solely determined by $ g^2/\lambda $, 
and  there is a strong resonance at the longest wavelengths for 
$ g^2/\lambda =2, 8, 18, \cdots $. 
Roughly speaking, the largest wave number in the efficient
resonant band is 
\begin{equation}
k_{max} = \left({g^2\over \lambda} \right)^{1/4}\sqrt{\lambda} \phi_0,
\label{kmax}
\end{equation}
An outstanding feature of conformal models is that the modes 
which are amplified by parametric resonance do not change  
by cosmic expansion. 
The Lam\'{e} equation for $\phi$ also has instability bands, 
but the growth rate is small compared with the typical 
one for $\chi$ and limited to roughly the Hubble scale \cite{quartic,pgreene}.

In our lattice simulations the evolution of the scale factor $ a $ is determined self-consistently
by solving Friedmann equation with the spatially averaged energy density.
In the simulation, $ \lambda $ is fixed to $9\times 10^{-14}$ 
appropriate to the COBE normalization.
We studied both $ g^2/\lambda =2$ and $50$.  
In both cases there is parametric resonance of $ \chi $ for $k=0$ mode
in the linear regime. In the former, the $\chi$ background is not strongly suppressed while in
the latter case the $\chi$ field is heavy and is strongly suppressed during inflation 
\cite{jedamzik,liddle,bassett2}. 

\subsubsection{Massive Inflaton Models}
We also considered massive inflaton models with 
the potential 
\begin{equation}
V(\phi, \chi)=\frac{m^2}{2} {\phi}^2+\frac{g^2}{2} {\phi}^2 {\chi}^2. 
\label{massive1}
\end{equation}
When the inflaton field oscillates around the 
potential minimum, 
the equation of state of the inflaton is dust on average. 
This means that amplitude of the background inflaton field decreases as
$ \propto a^{-3/2} $.  
Therefore it is convenient to introduce rescaled fields as
\begin{equation}
{\tilde \phi}=a^{3/2} \phi /{\phi}_0, \ \ {\tilde \chi} = a^{3/2} \chi /{\phi}_0, \label{massive2}
\end{equation}
where $ {\phi}_0 =0.193 m_{pl} $. Then, until the back reaction due to parametric resonance becomes 
efficient, the amplitude of the background $ {\tilde \phi} $ 
stays almost constant.
In terms of rescaled fields
the equations of motion are 
\begin{eqnarray}
{\ddot {\tilde \phi}} - a^{-2} \triangle {\tilde \phi} - \frac{3}{2} {\dot H} {\tilde \phi} - \frac{9}{4} 
H^2 {\tilde \phi} + m^2 {\tilde \phi} + g^2 a^{-3} {\tilde \phi} {\tilde \chi}^2 =0, \nonumber \\
{\ddot {\tilde \chi}} - a^{-2} \triangle {\tilde \chi} - \frac{3}{2} {\dot H} {\tilde \chi} - \frac{9}{4}
H^2 {\tilde \chi} + g^2 a^{-3} {\tilde \phi}^2 {\tilde \chi} =0. \nonumber \\ \label{massive4}
\end{eqnarray}
where $ \dot{} $ denotes the differentiation with respect to 
the proper time.
From these equations, in contrast to the conformal case, we see that the expansion of the universe 
affects the motion of scalar fields:
the wavelength of each comoving mode is redshifted 
and the effective coupling between $\tilde \phi$ and 
$\tilde \chi$ is decreased.

The linearized equations of Eqs. (\ref{massive4}) was 
extensively studied in~\cite{kofman2}. 
The equation for $ \chi $ approximately reduces to the so-called 
Mathieu equation and the evolution of $ \chi $ field shows 
broad resonance for 
\begin{equation}
q := \frac{g^2 {\phi}_0^2}{4 m^2} \gg 1. 
\label{massive5}
\end{equation}
Huge amplification of $\chi$ occurs at each time when 
amplitude of the inflaton field becomes zero due to violation of 
the adiabatic condition $\dot{\omega}/\omega^2 < 1$ for the $\chi$-field 
frequency, $\omega$.

When the expansion of the universe is taken into account 
the parametric resonance shows stochastic behavior 
because the phase of the $ \chi $ field is randomized. 
Despite the stochastic nature, the amplitude of $ \chi $ 
field grows exponentially on average. 
The maximum wave number in the resonance band is given by\cite{kofman2},
\begin{equation}
k_{max} \sim m q^{1/4}. 
 \label{con4}
\end{equation}
In our simulations we adopt $ m=10^{-6} m_{pl} $
indicated by the COBE normalization, 
and $ g=10^{-3} $ as a representative 
which realizes strong resonance. 
With this choice of parameters, 
we have $ q = 10^4$ at the beginning of preheating.   

\subsection{Initial Conditions}
Since the modes that our simulation covers 
are mostly on subhorizon scales, 
the initial conditions for the field fluctuations after 
inflation can be determined 
by the formula for the adiabatic vacuum. 
For example, for the $\chi$-field we have
\begin{equation}
\langle {\chi}_k {\chi}_{-k'} \rangle = 
  \frac{1}{2 {\omega}_k}\delta (k-k'),
 \label{ini5}
\end{equation}
where $ {\omega}_k = \sqrt{m^2_\chi+k^2} $ and 
\begin{equation}
m_\chi:=g\phi_0, 
\end{equation}
is the effective mass evaluated at the end of inflation. 
Recall that we have set $a=1$ at that time. 
Since we consider the cases with 
$m_\chi/k_{max}=(g^2/\lambda)^{1/4}$ greater than unity ( for $ g^2/\lambda=50 $), 
the power spectrum of $ \chi $ field is given by   
\begin{equation}
{\cal P}_{\chi}(k) \approx \frac{k^3}{4\pi g \phi_0^3 }, 
\end{equation}
for wavelengths relevant for parametric resonance. 

We also use Eq.~(\ref{ini5}) as the initial condition for 
fluctuations of $ \tilde\phi $, replacing $ m_\chi $ in 
$ {\omega}_k $ with 
\begin{equation}
 m_{\phi}:=\sqrt{3\lambda}\phi_0,
\end{equation}
in the conformal models and 
$m$ in the massive inflation models. 

In the conformal model with $ g^2/\lambda =2 $, 
$ \chi $ stays almost always massless during inflation. 
In this case the initial spectrum on the superhorizon size
becomes scale invariant \cite{bassett2,zibin}. 
However the precise initial power spectrum ( i.e. after inflation) of scalar fields on 
subhorizon scales is rather involved.
So here, for simplicity, we approximated $ {\cal P}_{\chi} $ as
\begin{equation}
{\cal P}_{\chi} = \frac{1}{4 {\pi}^2} (k^2+H^2), \ \ \ (a=1), \label{ini6}
\end{equation}
which is blue on small scales and flat on large scales, capturing the key features of
the spectrum. We will see that the precise form has no effect on the final results.

In our simulations, we compute classical dynamics 
taking the variance of these initial quantum fluctuations 
as if it were statistical variance, as standard \cite{latticeeasy}. 
During the early stage the evolution is in linear regime and amplification of the 
$\chi$ occupation number in the quantum picture 
is correctly described by the amplification of the perturbation amplitude
in classical dynamics \cite{son}. 
When the nonlinearity becomes important,  
the occupation number of modes relevant for resonance 
is far beyond unity. 
Hence, a classical treatment is justified at late epoch, too. 
Moreover as we will see later, our final conclusion is 
quite insensitive to initial conditions. 

\section{Numerical Results}
\subsection{Conformal Models}
\subsubsection{ $ g^2 / \lambda =50 $ case}

\begin{figure}[htbp]

\hspace{0.5cm}

  \includegraphics[width=7cm,height=16cm,clip]{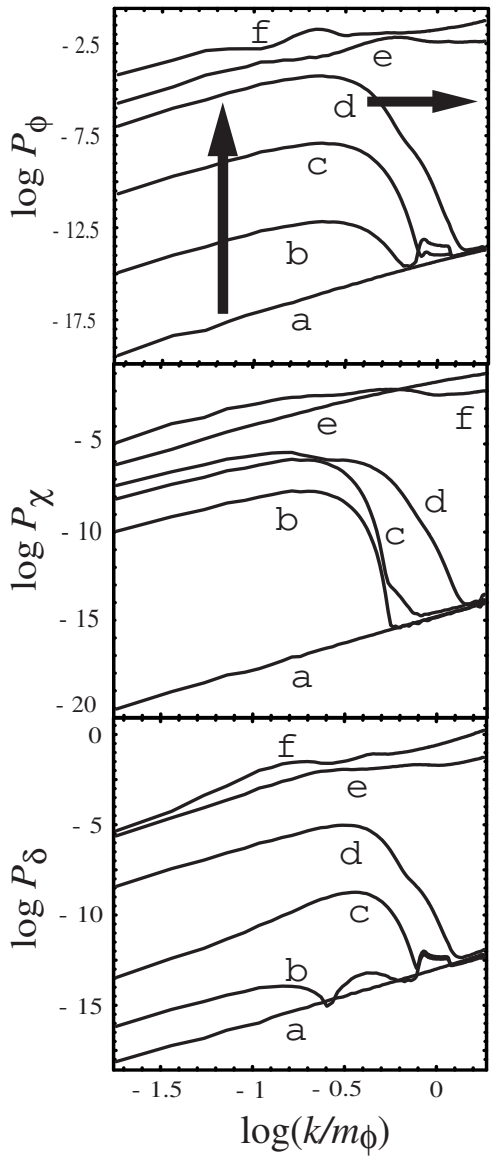}

\hspace{0.5cm}

\caption{Evolution of the power spectrum of scalar fields
and the total density perturbation for $ g^2/\lambda =50 $.
The horizontal axis is comoving wave number normalized by the (effective) 
inflaton mass at the end of inflation $ m_{\phi} $. 
Time flows vertically. Lines labeled as a, b, c, d, e and f are for times 
$ {\tilde \eta}=0, 50, 60, 70, 80, 100 $ respectively.
Thus during our range of simulation, the inflaton field 
oscillates about $ 100/7.4 \sim 14 $ times. 
An arrow pointing upward represents 
rapid increase of the amplitude of fluctuations 
of the inflaton field due to the initial parametric resonance, 
and the arrow pointing rightward shows rescattering: 
high energy particles are generated by mode-mode coupling and 
scattering of low energy particles.}
\label{g50}
\end{figure}

{}Fig. \ref{g50} shows the time evolution of the power spectra of the $\phi$
and $\chi$
fields and density perturbations until the backreaction shuts off the
parametric resonance. Let us first focus on the power spectra of 
$\phi$ and $\chi$ fields. Both the power spectra
are proportional to $ k^3 $ at the beginning. 
This is because the mass of each field is larger than the 
highest momentum resolved by simulation.
In the early stage of evolution linear perturbation is a good
approximation. 
The maximal characteristic (Floquet) exponent $ {\mu}_{max} $ for $ \phi $
is $ \sim 0.036 $, while that for $ \chi $ is $\sim 0.2 $. 
Hence, in this early stage perturbations of $\chi$-field grow
exponentially, but those of $\phi$-field almost stay constant. 
After a few oscillations of the inflaton field, the perturbations 
of $\phi$ suddenly start to grow. This can be understood as
follows. 
From Eq.~(\ref{model91}) , the equation of motion for 
$\tilde\varphi:=\tilde\phi-\langle\tilde\phi\rangle$ is 
\begin{equation}
\tilde \varphi'' -\lambda^{-1}\triangle \tilde\varphi + 
   3\langle\tilde\phi\rangle^2 \tilde\varphi  
   + 3 \langle\tilde\phi\rangle \tilde\varphi^2 
   + \tilde\varphi^3 + {g^2\over \lambda} 
    (\langle\tilde\phi\rangle +\tilde\varphi) \tilde\chi^2 =0.  
\label{result1}
\end{equation}
As $ \chi $ grows exponentially by the parametric resonance, the last 
term in Eq.~(\ref{result1}), 
$ \frac{g^2}{\lambda} \langle\tilde \phi\rangle \tilde\chi^2 $,  
exceeds the term $ 3 {\langle \tilde\phi \rangle}^2 \tilde\varphi $. 
At this stage, the linear approximation for $ \tilde\varphi $ 
breaks and $\tilde\varphi$ 
begins to grow proportional to $\tilde\chi^2$, whose  
characteristic exponent is $\sim 0.4$ \cite{kofman2}. This happens when $\tilde\chi$ 
exceeds a critical value,
\begin{equation}
 \tilde\chi_c \sim \left(\frac{g^2}{3\lambda}\right)^{-1/2} \tilde\varphi^{1/2}. 
\label{result2}
\end{equation}
At the largest wave number in the resonance band of $\chi$-field, 
the initial amplitude of $\tilde \varphi$ is given by 
$\approx \sqrt{k_{max}^3/4\pi\sqrt{\lambda}}$
$\approx(g^2/\lambda)^{3/8}\sqrt{\lambda/4\pi}$. 
Hence $\tilde\chi_c$ is estimated as
\begin{equation}
\tilde\chi_c\approx \left({g^2\over\lambda}\right)^{-1/16}\lambda^{1/4} \sim 10^{-3}.
\end{equation}
This rough estimate is consistent with the results of our lattice simulations. 

Exponential amplification due to parametric resonance 
still continues until the effect of backreaction becomes significant.
Parametric resonance ends 
when the second term in Eq.~(\ref{model1})
becomes equal to the first term, that is when $\tilde \chi$ is amplified
to 
\begin{equation}
 \tilde\chi_b 
    = \left(\frac{g^2}{\lambda}\right)^{-1/4}. 
\label{result4}
\end{equation}
The initial amplitude of $\tilde\chi$ 
at the shortest resonant mode given in $(\ref{kmax})$
is $\approx \sqrt{k_{max}^3/m_\chi} \approx {(g^2/\lambda) }^{1/8} \sqrt{\lambda} $. 
Approximating the evolution of $ {\tilde \chi} $ 
as $ {\tilde \chi} \propto e^{\mu \tilde \eta} $, 
the time at which parametric resonance ends, $\tilde\eta_f$,
is estimated by 
\begin{equation}
e^{\mu \tilde\eta_f} =  \left(\frac{g^2}{\lambda}\right)^{-3/8} 
   \lambda^{-1/2}. 
\label{result5}
\end{equation}
Solving Eq. (\ref{result5}), 
we have $ {\tilde \eta}_f \sim 70 $. This estimate is
consistent with the result 
$\tilde\eta_f \sim 80$ read from Fig.~\ref{g50}. 
Since $ {\tilde \eta}_f $ depends on the initial amplitude of $\tilde\chi$ 
logarithmically, $\tilde\eta_f $ does not depend so much on 
the parameter $ g^2/\lambda $ as long as models associated with 
strong resonance are concerned \cite{kofman2}.

By the time when the backreaction becomes important, 
the amplitude on smaller scales also increases due to the effect of 
rescattering. 

\begin{figure}[h]

\hspace{1cm}

  \includegraphics[width=7cm,clip]{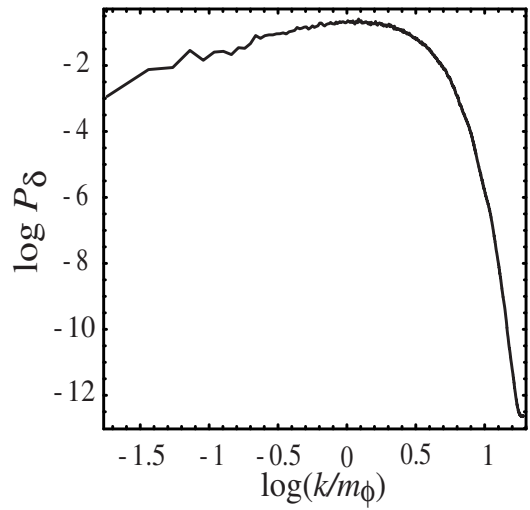}

\hspace{0.5cm}

\caption{The power spectrum of density perturbation after preheating for two-dimensional space.
We see that the slope of the power spectrum for small $ k $ is two, as expected.}
\label{2dim} 
\end{figure}

In three dimensions, at the time when the simulation ends, the shortest 
wavelength modes have the largest amplitude in the simulation. 
However, this seems to be an artifact due to lack of resolution. 
In the corresponding two dimensional simulations shown in
Fig.\ref{2dim}, we can also see the rescattering effect. 
In this case perturbations do not pile up near the 
shortest wavelength but peak at a finite value of $k$. This is indeed consistent 
with the picture that $\chi$ particles with 
very large kinetic energy $k \rightarrow \infty$, are not produced.

Let us now focus on the power spectrum of density perturbations.
The energy density $\rho$ is given by 
\begin{eqnarray}
a^4\lambda^{-1}\rho = 
 \frac{1}{2} { ( {\tilde \phi}' - {\tilde {\cal H}} {\tilde \phi} ) }^2 
   + \frac{1}{2} (\tilde\chi'-\tilde{\cal H}\tilde\chi)^2 
    + \frac{1}{2\lambda} (\vec\nabla \tilde\phi)^2 \nonumber \\ 
+\frac{1}{2\lambda} (\vec\nabla \tilde\chi)^2 + \frac{1}{4}
 \tilde\phi^4+\frac{g^2}{2\lambda} \tilde\phi^2 \tilde\chi^2, 
\label{result6}
\end{eqnarray}
where ${\tilde {\cal H}}:=\partial_{\tilde\eta}\log a$. 
{}From this equation, 
the density perturbations to first order are 
\begin{equation}
a^4\lambda^{-1} \delta \rho \approx (\langle\tilde\phi\rangle' - 
 \tilde{\cal H}\langle\tilde\phi\rangle) 
   ( \varphi'-\tilde{\cal H}\varphi) + \langle\tilde\phi\rangle^3
   \varphi. 
\label{result7} 
\end{equation}
As we have already mentioned, 
amplitude of $\varphi$ is almost constant in the linear 
perturbation regime.  
Therefore in this regime 
amplitude of density perturbations does not grow. 
This behavior of $\delta\rho$ can be observed in Fig.~\ref{g50}. 
When the amplitude of $\tilde\chi$
reaches $\tilde\chi_c$, the terms second order in field 
perturbations start to contribute to $\delta\rho$. 
Then, density perturbations begin to grow rapidly. 
After the exponential increase of $\delta\rho$, the
amplification of density perturbations stops when the growth of field 
perturbations terminates due to back reaction to the oscillation of 
$\langle\tilde\phi\rangle$. 

{}From Fig.~\ref{g50}, we find that the slopes of
power spectrum of resultant density perturbations on
large scales are all equal to three. 














As we shall see below, this result 
does not change even if we artificially amplify the initial fluctuations
of fields on large scales, which leads to the rule of thumb that after
preheating correlation of perturbations on large scales 
disappears rather independently of initial power spectrum. 




We can give a simple interpretation to this result. 
What we assume is that the resultant density perturbations 
have a typical scale $r$, and correlations on larger length scales 
are strongly suppressed. In such a situation, the integral 
\begin{equation}
 {1\over k^D}{\cal P}_{\delta}(k)
    \propto \int d^D x\,
        \langle\delta(0) \delta({\vec x})\rangle 
          e^{i\vec k\vec x},  
\end{equation}
is dominated by a small region with $|\vec x|\alt r$, 
where $D$ is the number of spatial dimensions, 
which is 3 in our simulation. For 
long wavelength modes with $k\ll r^{-1}$, 
$e^{i\vec k\vec x}$ can be approximated by unity. Thus we have 
\begin{equation}
{\cal P}_{\delta}(k) \propto k^D,
\label{result11}
\end{equation}
We also performed 2 dimensional lattice simulations, 
and found the power spectrum proportional to $k^2$ on large scales as predicted; see
(Fig.~\ref{2dim}). 

One may think that we have obtained  
this result because of the initial blue spectrum of $\chi$-field.  
Since the characteristic exponent of the parametric amplification 
is almost the same for all modes with 
$ k/m_\phi < { (g^2/\lambda) }^{1/4}$, 
the parametric resonance will end due 
to the backreaction from the mode of 
the shortest resonance scale, at which the initial amplitude 
is the largest among the modes in resonance. 
At that time perturbations of $\chi$-filed still remain 
small on large scales (where by large scale we here mean around the horizon size and larger).

\begin{figure}[h]

\hspace{1cm}

  \includegraphics[width=7cm,clip]{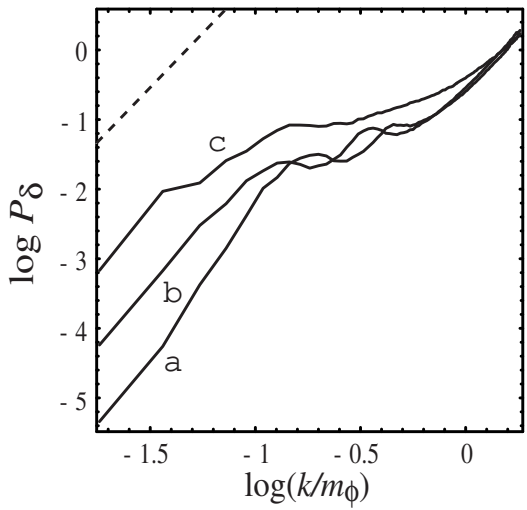}

\hspace{0.5cm}

\caption{Power spectrum of the total density perturbation after preheating for three cases in three dimensions.
Lines labeled as a, b and c correspond to $ g^2/\lambda=50 $, $ g^2/\lambda=50 $ with
scale-invariant power spectrum and $ g^2/\lambda=2 $ respectively. 
We see that the slope of the power spectrum for small $ k $ is universal with a value of three.
The dashed line has $ k^3 $ power and crosses the threshold at the horizon scale.
All three lines are lie significantly under the dashed line, showing that PBHs are not overproduced in 
these cases. The peak of the spectrum 
and subsquent decay are not resolved by our three dimensional simulations but are resolved in two dimensions.
}
\label{final} 
\end{figure}

Here we show that the initial blue spectrum is not 
a necessary condition for suppression on large scales. 
For this purpose, we performed the same simulation but with the scale invariant 
initial spectrum (${\cal P}_\chi$=constant, where the power spectrum is initially
amplified on large scales).
Fig. \ref{final} shows $ {\cal P}_{\delta} $ after preheating in this case. 

From this figure, we see that at the end of preheating
perturbations on horizon scales are suppressed with slope 3, which is
the same as in the case with the initial blue spectrum. 
This result indicates that in general parametric resonance causes loss
of correlation between density fluctuations beyond a typical length


scale and energy is efficiently cascaded to shorter wavelengths by rescattering.

In order to estimate the production rate of PBHs,  
we have to compute $\delta\rho/\rho$ smoothed over the horizon size. 
The horizon size when parametric resonance ends corresponds 
to $k=aH\approx 5 \times 10^{-3}m_{\phi}$, 
which is not covered in our three-dimensional simulations. 



However, since there is no typical length scale before the horizon scale, 
it will be natural to expect that 
one can extrapolate the power spectrum to horizon size 
assuming the slope of the power spectrum of density
perturbations is $D$. 
The result of 2 dimensional simulations (Fig.~\ref{2dim}) 
also support this extrapolation. 
With the aid of this extrapolation, 
the amplitude of density contrast at the horizon size 
$ {\delta}_H $ can be estimated from Fig.~\ref{g50} as
\begin{equation}
{\delta}_H \sim 3 \times 10^{-4}, \label{result12}
\end{equation}
which is an order of magnitude smaller than the threshold for PBH overproduction (0.03). 
Therefore we conclude it is unlikely that PBHs are overproduced 
by the parametric resonance in the case with $g^2/\lambda = 50 $.

\subsubsection{$ g^2 / \lambda =2 $ case}
We also performed lattice simulations for $ g^2/\lambda =2 $. In this
case, the power spectrum of $ \chi $ at the end of inflation is
flat for $ k < H_0 $, where $ H_0 $ is the Hubble parameter
at the end of inflation. Therefore, the modes whose wave length is much larger
than the horizon size is not suppressed in this case, which is different
from the case 1 $ g^2/\lambda \gg 1 $ \cite{bassett2}.  Taking into account the fact that
there is a strong resonance at small $ k $ for $ g^2/\lambda =2 $ ,
there is a possibility of PBHs overproduction in this
case~\cite{bassett3,agreen}. 

$ {\cal P}_{\delta} $ for $ g^2/ \lambda =2 $ is shown in Fig.~\ref{final}.
As before,  the power spectrum of $\delta$ after parametric resonance is suppressed 
and its slope is three around the horizon scale, implying that large scale perturbations are uncorrelated.
Therefore in the same way as discussed in the case with $g^2/\lambda=50$, production of PBHs caused 
by parametric resonance will not be efficient enough to exceed 
the astrophysical bounds. 

\subsection{Massive Inflaton Models}
\begin{figure}[h]

\hspace{1cm}

  \includegraphics[width=7cm,clip]{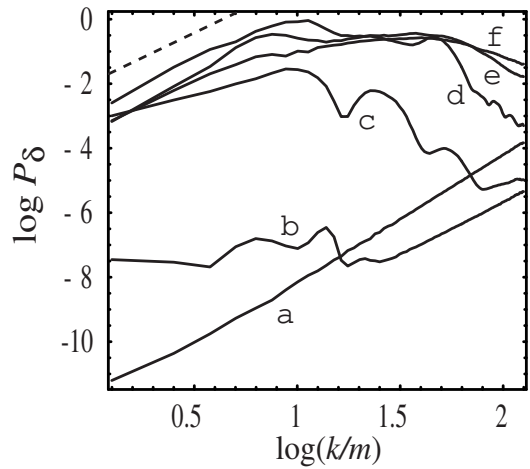}

\hspace{0.5cm}

\caption{Evolution of the power spectrum of $ \delta $ for the
massive inflaton model. Lines labeld as a, b, c, d, e and f
correspond to $ mt =0, 50, 60, 70, 100, 120 $ respectively.
The dashed line represents the threshold for PBH over-production which lies well above any of the curves. }
\label{massrho} 

\end{figure}

Fig.~\ref{massrho} shows the evolution of the power spectrum during preheating 
for a massive inflaton model. 
From this figure we find that the power spectrum after preheating 
is $\propto k^3$ on horizon scales as in the case of conformal models.

Here we cannot directly use the criterion for the PBH production discussed in
section II, because the universe is not radiation dominated but dust on average.
Due to the difference of the equation of state, 
the condition that density perturbations
collapse to form a black hole differs from the one 
in the radiation dominated universe.
In~\cite{carr3}, the critical density $ {\delta}_c $ for the equation 
of state 

\begin{equation}
P = \epsilon \rho, 
\label{remass6}
\end{equation}

depends on $\epsilon$. 
Here we assume that the instability of the density perturbation of scalar
fields in massive inflaton models is similar to the fluid case with the same 
effective equation of state. 

\begin{figure}[h]

\hspace{1cm}

  \includegraphics[width=7cm,clip]{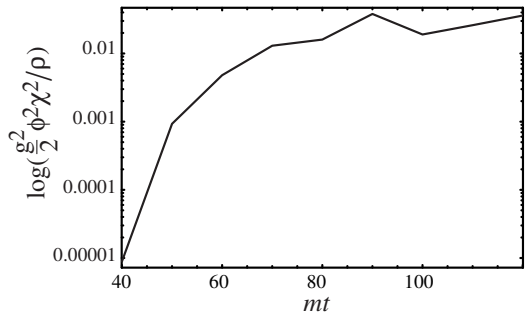}

\hspace{0.5cm}

\caption{Time evolution of the ratio of the interaction energy $ \frac{g^2}{2} {\phi}^2 {\chi}^2 $ 
to the total energy density $ \rho $. Time is normalized in $ 1/m $. We see that $ \epsilon $ after
preheating is about $ 3 \times 10^{-2} $, i.e. the total system behaves approximated as dust.}
\label{EOS}
\end{figure}

The energy density and pressure in massive inflaton
models are
\begin{eqnarray}
 \rho = T+U, \qquad P=T-U,
\label{remass1}
\end{eqnarray}
with
\begin{eqnarray}
T &= &\frac{1}{2} {\dot {\phi}}^2 + \frac{1}{2} {\dot {\chi}}^2,
\cr
U &= & \frac{1}{2} {(\frac{1}{a} \vec{\nabla} \phi)}^2
+\frac{1}{2} {( \frac{1}{a} \vec{\nabla} \chi)}^2 + \frac{m^2}{2} {\phi}^2 +
\frac{g^2}{2} {\phi}^2 {\chi}^2.
\end{eqnarray}
Using the equations of motion for the scalar fields, we 
can show the relation,
\begin{equation}
\langle T \rangle = \langle U \rangle + \frac{g^2}{2} 
\langle {\phi}^2 {\chi}^2 \rangle, 
\label{remass3}
\end{equation}
where $\langle \cdots\rangle$ denotes a long time average with 
the weight $a^3\,dt$.  
Hence we can estimate $\epsilon$ by 
\begin{equation}
\epsilon \approx 
\frac{g^2}{2} \langle {\phi}^2 {\chi}^2 \rangle /\langle\rho\rangle. 
\label{remass5}
\end{equation}
Fig.~\ref{EOS} shows the time evolution of this quantity during preheating. 

At the end of preheating 
$\epsilon$ becomes as large as $ \sim 3 \times 10^{-2}$. 
Therefore the ratio of pressure to energy density after preheating in
massive inflaton model is ten times smaller than that of 
the radiation dominated universe. 
Hence, the upper limit on 
$ \sigma $ in the case of massive inflaton model will be reduced to
about 
$ 3\times 10^{-3} $. 
On the other hand, from Fig.~\ref{massrho}, the value
of the power spectrum of the density perturbation at horizon size 
($k\approx 1\times 10^{-1}m $) can be estimated
as
\begin{equation}
{\delta}_H \sim 5 \times 10^{-4}, \label{remass7}
\end{equation}
where we have used the simulations to conclude that the power spectrum is proportional to $k^3$ 
for small $ k $. This is smaller than the threshold $ \sim 3\times 10^{-3}$ again by about one order of magnitude. 
Therefore we conclude that PBHs will not be
overproduced in massive inflaton model, either, despite the fact that $\sigma$ is significantly enhanced 
by resonance on horizon scales.

\begin{figure}[h]

\hspace{1cm}

  \includegraphics[width=7cm,clip]{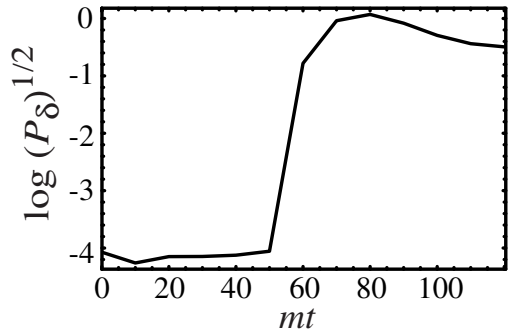}

\hspace{0.5cm}

\caption{Time evolution of $ \sqrt{ {\cal P}_{\delta} } $ at $ k=k_{max}=q^{1/4} m $.
From this, we see that backreaction terminates the parametric amplification
at around $ mt = 70 $.}

\label{backreaction}

\end{figure}

\section{Gaussianity}
In this section, we discuss the probablity 
distribution of the amplitude of density perturbations 
at the end of preheating.
In the preceding sections we estimated 
the abundance of PBHs by assuming that the probability distribution of
density perturbations 
at the horizon size is Gaussian. 
If the tale of distribution, 
which is relevant for PBH formation, had a non-Gaussian tail, the resultant astrophysical 
constraints would be significantly altered and hence it is a crucial assumption to test. 
Further, since rescattering ($\propto \delta \chi^2$) is crucial, one might expect chi-squared
corrections to be important.

\begin{figure}[h]

\hspace{1cm}

  \includegraphics[width=7cm,clip]{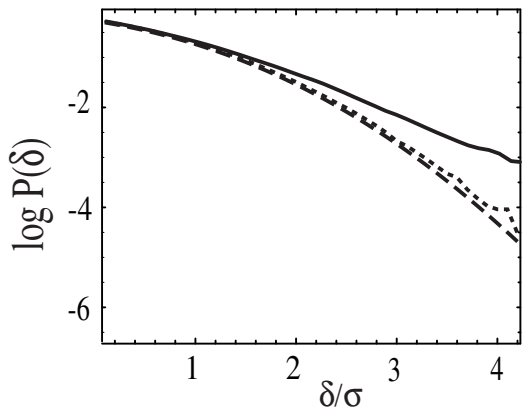}

\hspace{0.5cm}

\caption{The probability distribution of density fluctuations smoothed over the scale $ m_{\phi} $ (the Compton wavelength 
of the inflaton) in the conformal model.
The dotted line is the spectrum at initial time and black line is the spectrum at the end of preheating.
The dashed line is an appropriately scaled Gaussian distribution. 
At the end of preheating, the distribution of large fluctuations is significantly 
more amplified than the Gaussian prediction.}
\label{Sgauss}

\end{figure}

We first show the distribution of density fluctuation smoothed over the shortest
resonant scale in a conformal model. The result is shown in Fig. \ref{Sgauss}.
We can see that the distribution does not trace a Gaussian distribution 
at the end of preheating, while it does at the 
initial stage. In particular, the probability of large amplitude of 
perturbations is enhanced through preheating. 
Interestingly, the late time distribution 
looks like exponential. 

This results can be understood as follows. 
At initial stage where the linear approximation is valid, 
density perturbation is just a superposition of Gaussian distributions. 
Hence, the probability distribution is Gaussian.
As perturbations grow, the terms quadratic in field 
perturbations start to contribute to $\rho$.
The probablity distribution in such situation will 
be mimiced by a product of two Gassian random 
variables $ x $ and $ y $. 
The probability
distribution of $ z=xy $ is given by 
\begin{eqnarray}
P(z) & = & \frac{1}{2 \pi {\sigma}_1 {\sigma}_2} \int \frac{dx}{x} e^{ - \frac{x^2}{2} -\frac{z^2}{2 {\sigma}^2_1 
{\sigma}^2_2} \frac{1}{x^2}}\cr
    & \sim & \frac{1}{2 \sqrt{2 \pi {\sigma}_1 {\sigma}_2 z} } e^{- \frac{z}{
{\sigma}_1 {\sigma}_2 }}, 
\label{gauss2}
\end{eqnarray}
where $\sigma_x^2$ and $\sigma_y^2$ are variances $x$ and $y$, respectively. 
Here in the last step 
we used the steepest decent method assuming 
$z$ is much larger than $ {\sigma}_1 {\sigma}_2 $.  
In this manner one can reproduce a pure 
exponential distribution for large values of $ z $. 

\begin{figure}[h]

\hspace{1cm}

  \includegraphics[width=7cm,clip]{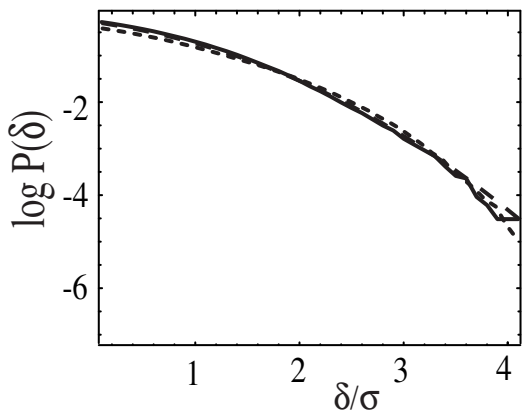}

\hspace{0.5cm}

\caption{The probability distribution of density fluctuation smoothed over the scale 
$ 50 \times m_{\phi} $ in the conformal model. The dotted line is the distribution at the initial 
time and the black one is at the end of preheating. Contrary to Fig. \ref{Sgauss}, 
the distribution is still Gaussian even at the end of preheating. Hence, on the horizon scale 
relevant to PBH production the Gaussian assumption is a good approximation.}
\label{Lgauss}

\end{figure}

Next we show the distribution of density perturbations 
averaged over a large scale. The result is shown in Fig.~\ref{Lgauss}. 
In this case the distribution is almost Gaussian even at the end of preheating.
This result can be interpreted as follows. As we discussed in 
section V, the density perturbations lose correlation on 
scales much larger than the shortest wavelength in the resonance band. 
Hence, the average over a large length scale $L$ behaves as a sum of 
a large number of independent random variables 
of $O\left((L k_{max}/a)^3\right)$.   
Therefore its distribution is guaranteed to be close to Gaussian 
by the central limit theorem, which is consistent with numerical
the results. Significant deviations from Gaussianity are not expected unless 
the amplitude is as about $(L k_{max}/a)^3$ times larger than the standard 
deviation. Since the horizon 
size at the end of preheating is much larger than 
the shortest wavelength in the resonance band, 

the required amplitude is extremely large, 
and hence the probability of finding it 
is completely negligible. 




Hence, non-Gaussianity cannot affect the estimate of 
the PBH formation rate. 

\section{The effect of metric perturbations}
Finally we briefly discuss the role of gravitational interactions (metric perturbations) 
which might have effects on PBH formation.  We have 
neglected them throughout this paper but  there is a good reason why one expects that this is a 
good approximation.
The time scale for gravitational collapse is at most  
free fall time. Unless density perturbations significantly 
exceed $O(1)$, this time scale is identical to the Hubble time scale.
On the other hand parametric resonance undergoes with the time scale 
determined by the effective mass of the inflation, which is in general 
shorter than the Hubble time scale. 
Moreover, in the expanding universe gravitational instability 
does not grow exponentially, while parametric resonance drives 
exponential growth of perturbations. 
Hence, we expect gravitational interactions play a subdominant role 
at the stage of preheating, although later on a longer time scale 
gravitational collapse may proceed in cases where the  
effective pressure happens to be very small (such as in the massive model). 

In the treatment neglecting the gravitational interaction,  
there arises another subtlety related to the gauge. 
We discussed the amplitude of density perturbations at 
the horizon size, but it is a gauge dependent quantity. 
Hence, strictly speaking, it is incorrect to quote the criterion 
for the PBH formation stated in terms of density perturbations 
in comoving gauge. 
Moreover, it is more suitable to use the amplitude of 
metric perturbations rather than density contrast 
\cite{niemeyer,shibata}. 

In the present context, 
this is mainly because the power spectrum of density perturbation 
has a strong $k$-dependence, $\propto k^3 $, 
which means that probability of PBH formation is very sensitive 
to the choice of the horizon size~\cite{bassett3}. 
In contrast, since 
the metric perturbations well inside the horizon 
is characterized by the Newton potential, 
the power spectrum of metric perturbations will 
be proportional to $k^{-1}$. 
On the other hand, it was shown in~\cite{tanaka} that the
curvature perturbation on the constant energy density hypersurfaces 
$\zeta $ (Bardeen parameter) behaves like $\propto k^3$
on super horizon scales after preheating 
by using the separate universe 
approach~\cite{stewart,taruya,KH,TS,wands}. 

Therefore, say, the curvature perturbation $\zeta$
will have a peak near at the horizon size, 
and we will be able to obtain an unambiguous upper limit 
on the abundance of PBHs produced by preheating. 
From the above consideration, 
including metric perturbation in the evolution of 
scalar fields is an interesting issue which we leave for
future work.

\section{Summary and discussion}
We have studied the formation of black holes during preheating after 
inflation. Preheating provides a challenging framework to illucidate various complex 
physical processes such as non-equilibrium, non-perturbative field theory in 
expanding backgrounds. In addition, since preheating is generic in some regions of 
parameter space for many inflationary models, the issue of whether primordial black holes (PBH) 
are overproduced is an important one. 

To address this we have performed two and three dimensional lattice simulations 
of several different inflaton potentials (conformal and massive) and parameter regions. These simulations automatically incorporate all non-linear effects such as 
backreaction and rescattering of fields. We found no evidence for over-production of PBH, in 
contrast to earlier expectations. In addition we found that although highly non-Gaussian on very small scale,
the spectrum of density perturbations is effectively Gaussain on the horizon (i.e. Hubble) scale.  

Our results can be understood simply. For all cases, we found that when the amplitude of 
density perturbations at about 
the shortest wavelength in the resonant band becomes of order unity,  the growth due to parametric resonance terminates due
to backreaction.  At the end of preheating the final density spectrum is universal, with a
blue power spectrum, $ \propto k^D $ (D = 2,3) on horizon scales, depending on whether the simulation was two ($D=2$) 
or three ($D=3$) dimensional.  Since the peak of the spectrum is scales significantly shorter than the horizon scale, the amplitude of the
density perturbation at the horizon scale extrapolated from the peak is typically about an order of magnitude  below the threshold for PBH overproduction.

We gave an explanation of this universality in the slope of the final power spectrum 
on large scales as a result of loss of coherence 
due to parametric resonance. 

These results argue for the view that generically PBHs will not be produced so much as 
to violate astrophysical constraints, even in the case of strong preheating.  
This result also applies to the cases with other models and  parameters regions 
if our interpretation of the power spectrum on large scales is 
universally correct. 

The result obtained in this paper 
is different from the claim by Green and Malik~\cite{agreen}, where
they found that PBHs are likely to be overproduced. 
They estimated the time when backreaction becomes 
significant as well as the time when the amplitude of 
density perturbations exceeds the threshold separately
based on linear approximation. 
Comparison of these two times was used to give a criteria 
for PBH formation. However, for example, in Ref.~\cite{kofman2} 
the time when the backreaction becomes significant in the case 
with $m=10^{-6} m_{pl}$ and $ g=10^{-3} $ is estimated to be $ \sim 90 $,
which is slightly 
later than our numerical result (See Fig.~\ref{backreaction}). 
Since the growth of perturbation amplitude is exponential, 
a small error in the back reaction time can lead to wrong conclusions.
In calculating the abundance of produced PBHs, 
only $ 10 \sim 20 $ percent
error of backreaction time can lead to the opposite conclusion.
By contrast, our conclusion is based 
on self-consistent simulations and rather robust qualitative observations.  
There is no delicate comparison of different time scales. 

In this paper, we considered standard slow roll inflation. 
In such cases, preheating occurs at rather high energies. 
Therefore the mass of PBHs formed in the present context 
is too small to avoid evaporation before the Big Bang Nucleosynthesis. 
Hence, those PBHs are not subject to any observational
constraint even if PBHs are produced abundantly, 
unless PBHs leave Planck mass relics~\cite{barrow}. 
(Hence we have assumed that PBH will leave Planck
mass relic throughout this paper. )
Even if relics are formed, the 
constraint on the mass fraction of produced black holes 
is $ \beta < 10^{-20} $~\cite{carr5},  
we can therefore conclude that 
production of PBHs by preheating does not give a serious 
constraint on such simple models of preheating as discussed 
in this paper. 

However, our qualitative results will also apply for preheating at 
lower energies. Let us consider the possibility
of making more massive black holes by preheating, 
say, in the case of massive inflaton models. 
Now we consider to vary the inflaton mass $m$ and 
the value of $\phi$-filed at the beginning of preheating $\phi_0$ 
within $\phi_0 \alt m_{pl}$
(As an example of realizing small value of $\phi_0$, we can consider 
the hybrid inflation model~\cite{linde}.).  
The key quantity is the ratio of $ k_{max} $ given in (\ref{con4}) to
the Hubble parameter, which is estimated as
\begin{equation}
\frac{k_{max}}{H} \sim \frac{m_{pl}}{\phi} q^{1/4}. 
\label{con5}
\end{equation}
This ratio is independent of $m$ and is the smallest for $\phi\approx m_{pl}$. 
Then the same estimate given in (\ref{remass7}) applies 
as an upper bound for $\delta_H$. On the other hand, lowering $\beta$ 
down to $10^{-30}$ only change the upper bound on $\sigma$ 
from 0.031 to 0.026. 
Thus we can say that overproduction of more massive PBHs due to
parametric resonance is also unlikely. 

We have also confirmed Gaussianity of 
the probability distribution of density perturbations at the 
horizon scale, which is assumed in the estimate of the production rate 
of PBHs.  
The appearance of Gaussianity is in accordance with the interpretation 
of the spectrum on large scales. 
If perturbations become uncorrelated beyond the shortest resonant scale,  
perturbations at the horizon scale are given by average
of many statistically independent random variables. 
Thus Gaussianity  naturally follows from the central limit
theorem. 

\acknowledgements

We thank Naoshi Sugiyama, Shugo Michikoshi, Motoyuki Saijo, Takashi Nakamura and Misao Sasaki for
useful comments. HK is supported by the JSPS. This work is supported in part by Monbukagakusho
Grant-in-Aid Nos. 16740141 and 14047212.

\end{document}